\documentclass{iau} 
\usepackage{graphicx}

\title[G2 as a diffuse gas cloud] %% give here short title %%
{3D AMR simulations of the evolution of the diffuse gas cloud G2 in the Galactic Centre}

\author[M.~Schartmann et al.]   %% give here short author list %%
{M. Schartmann$^{1,2,3}$, A. Ballone$^{1,2}$, A. Burkert$^{1,2,4}$, S. Gillessen$^2$, 
 R. Genzel$^2$, O. Pfuhl$^2$, F. Eisenhauer$^2$, P.M. Plewa$^2$, T. Ott$^2$, 
 E.M. George$^2$ \and M. Habibi$^2$}

\affiliation{
$^1$ University Observatory Munich, Scheinerstra\ss e 1, D-81679 M\"unchen, Germany \\[\affilskip]
$^2$ Max-Planck-Institute for extraterrestrial Physics, Postfach 1312, Giessenbachstr., D-85741 Garching, Germany \\[\affilskip]
$^3$ Centre for Astrophysics and Supercomputing, Swinburne University
of Technology, P.O. Box 218, Hawthorn, Victoria 3122, Australia \\
email: {\tt mschartmann@swin.edu.au}
 \\[\affilskip]
$^4$ Max-Planck-Fellow\\[\affilskip] 
 }

\pubyear{2016}
\volume{322}  %% insert here IAU Symposium No.
\jname{The Multi-Messenger Astrophysics of the Galactic Centre}
\editors{S.~Longmore, G.~Bicknell \& R.~Crocker, eds.}
\begin{document}

\maketitle

\begin{abstract}
With the help of 3D AMR hydrodynamical simulations we aim at understanding G2's nature, 
recent evolution and fate in the coming years. By exploring the possible parameter space 
of the diffuse cloud scenario, we find that 
a starting point within the disc of young stars is favoured by the observations,
which may hint at G2 being the result of stellar wind interactions. 
\keywords{accretion, black hole physics, Galaxy: center, hydrodynamics, ISM: clouds, ISM: evolution}
\end{abstract}

\firstsection % if your document starts with a section,
\section{Observations}

\cite[Gillessen et al. (2012)]{Gillessen_12} discovered a fast moving object within the 
range of the S-star cluster close to Sgr A*. VLT NACO images show the object in L'-band, 
but not in K-band, 
indicating that it is a dusty, ionised gas cloud. 
The spatially resolved ionised gas emission (especially in the Brackett-$\gamma$ line) monitored  
with the SINFONI instrument allows to accurately 
constrain the orbit around Sgr A* to be highly eccentric (e=0.98) with a peri-centre 
passage at a distance of 2400 Schwarzschild radii. The cloud has a mass of 3 earth masses 
and an orbital period of 400 years. The tidal interaction with the central black hole is 
clearly visible in the developing gradients in observed position-velocity (PV) diagrams
\cite[(Gillessen et al., 2013a, 2013b)]{Gillessen_13a,Gillessen_13b}. 
An overlay of the PV diagrams of the years 2004 to 2014 shows that 
G2, a second cloud G1 and the {\it tail} might form a stream of gas
\cite[(Pfuhl et al., 2015)]{Pfuhl_15}.

\section{Simulated evolution of a diffuse cloud}
We employ 3D AMR hydro-dynamical simulations using the {\sc PLUTO} code in order to follow the 
evolution of a diffuse gas cloud which starts initially in pressure equilibrium with an 
ADAF-like atmosphere on the observed orbit. Fig.~1 shows that the spherical cloud is 
initially compressed due to the radially increasing atmospheric pressure and ram 
pressure interaction (Fig.~1a). Close to peri-centre, the massive black hole leads to 
tidal stretching (Fig.~1b)
and after the peri-centre passage, the cloud develops a nozzle-like accretion stream
feeding gas towards the centre (Fig.~1c) with small infall rates. However, despite the 
seemingly disruption of the cloud, the (normalised) 
Brackett-$\gamma$ emission on the sky closely follows the observed orbit, even after  
the cloud's peri-centre passage.

\section{Comparison to observations and conclusions}

By directly comparing our simulated position-velocity (PV) diagrams for various starting 
times (in pressure equilibrium) along the orbit (upper row in Fig.~2) with the observed 
ones (residuals in second row in Fig.~2), we find better correspondence for models with 
a starting time close to apo-centre of the best-fit G2 orbit, as inferred from a 
reduced $\chi^2$ analysis. This might indicate a formation scenario due to stellar
wind interaction within the disc of young stars \cite[(Calder\'{o}n et al., 2016)]{Calderon_16} or G2's origin from a 
gas streamer. Brackett-$\gamma$ light curves for these simulations show a (partly physical) 
mixing plateau and are roughly in agreement with the data. Similarly good overall 
agreement is found for the {\it Compact Source Scenario} (see conference contribution by 
Alessandro Ballone). However, in the latter scenario, the source leaves G2 behind in 
approximately 5 years from now and forms a separate cloud, allowing us to distinguish the two scenarios.

\begin{figure}
% \vspace*{-2.0 cm}
\begin{center}
 \includegraphics[width=0.8\textwidth]{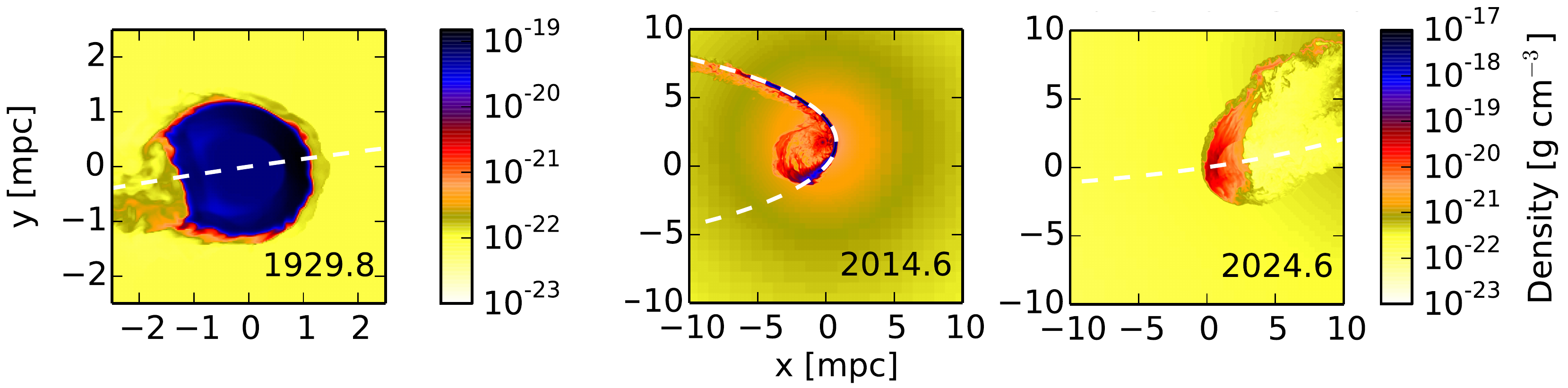} 
% \includegraphics[width=0.4\textwidth]{dens_slice_zoom_series_follow_orbit.pdf} 
% \vspace*{-1.0 cm}
 \caption{Density distributions showing the evolution from a spherical cloud to a thin
 filamentary structure: (a) compression due to increasing ambient pressure and ram pressure,
(b) tidal interaction and (c) accretion towards Sgr~A* in a nozzle-like accretion stream.
 Adapted from \cite[Schartmann et al. (2015)]{Schartmann_15}.}
   \label{fig1}
\end{center}
\end{figure}

 \begin{figure}
% \vspace*{-2.0 cm}
\begin{center}
 \includegraphics[width=0.75\textwidth]{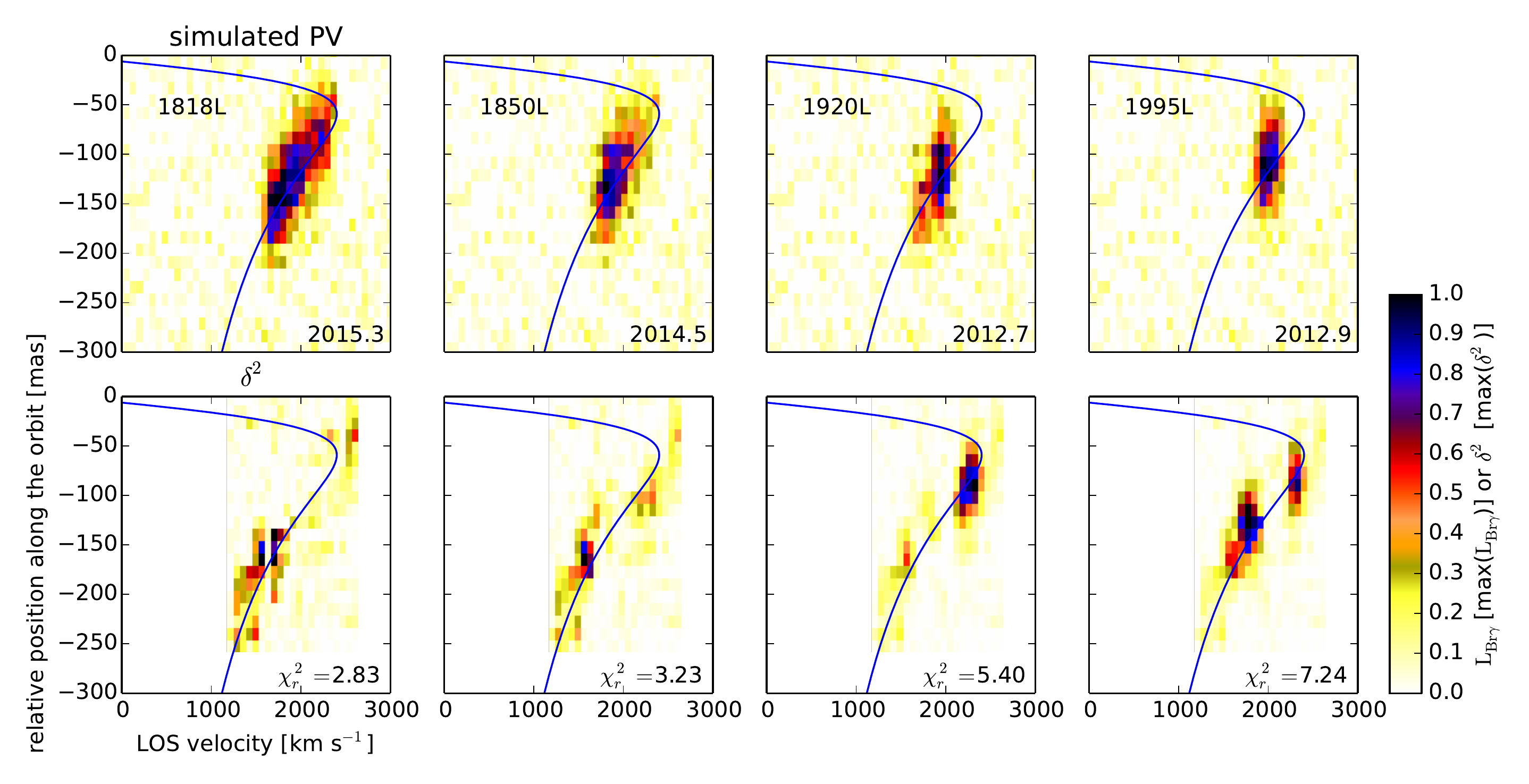} 
% \vspace*{-1.0 cm}
 \caption{Simulated PV diagrams (upper row) for various starting times of the clouds
 (as annotated in the upper left corner) and the normalized residual arrays after directly comparing them to the 2012 
 observed PV diagrams (lower row). The blue curve corresponds to G2's nominal orbit. A starting point 
 of the clouds close to peri-centre gives the best adaptation of the data. Taken from 
 \cite[Schartmann et al. (2015)]{Schartmann_15}.}
   \label{fig2}
\end{center}
\end{figure}


\begin{thebibliography}{}

\bibitem[{{Calder{\'o}n} {et~al.}}] {Calderon_16}
    {Calder{\'o}n}, D., {Ballone}, A., {Cuadra}, J., {et~al.} 2016, \textit{MNRAS}, 455, 4388

\bibitem[{{Gillessen} {et~al.}(2012){Gillessen}, {Genzel}, {Fritz}, {Quataert},
  {Alig}, {Burkert}, {Cuadra}, {Eisenhauer}, {Pfuhl}, {Dodds-Eden}, {Gammie},
  \& {Ott}}]{Gillessen_12}
{Gillessen}, S., {Genzel}, R., {Fritz}, T.~K., {et~al.} 2012, \textit{Nature}, 481, 51

\bibitem[{{Gillessen} {et~al.}(2013a){Gillessen}, {Genzel},
  {Fritz}, {Eisenhauer}, {Pfuhl}, {Ott}, {Cuadra}, {Schartmann}, \&
  {Burkert}}]{Gillessen_13a}
---. 2013a, \textit{ApJ}, 763, 78

\bibitem[{{Gillessen} {et~al.}(2013b){Gillessen}, {Genzel},
  {Fritz}, {Eisenhauer}, {Pfuhl}, {Ott}, {Schartmann}, {Ballone}, \&
  {Burkert}}]{Gillessen_13b}
---. 2013b, \textit{ApJ}, 774, 44

\bibitem[{{Pfuhl} {et~al.}(2015){Pfuhl}, {Gillessen}, {Eisenhauer}, {Genzel},
  {Plewa}, {Ott}, {Ballone}, {Schartmann}, {Burkert}, {Fritz}, {Sari},
  {Steinberg}, \& {Madigan}}]{Pfuhl_15}
{Pfuhl}, O., {Gillessen}, S., {Eisenhauer}, F., {et~al.} 2015, \textit{ApJ}, 798, 111

\bibitem[{{Schartmann} {et~al.}(2015)}]{Schartmann_15}
{Schartmann}, M., {Ballone}, A., {Burkert}, A.,
   {et~al.} 2015, \textit{ApJ}, 811, 155 

\end{thebibliography}
\end{document}